\documentclass{aa}
\usepackage{graphicx}
\usepackage{longtable}
\usepackage{txfonts} 

\begin{document}
\renewcommand{\topfraction}{0.85}
\renewcommand{\bottomfraction}{0.7}
\renewcommand{\textfraction}{0.15}
\renewcommand{\floatpagefraction}{0.66}

\title{Very high energy gamma rays from the direction of Sagittarius A$^*$}

\author{F. Aharonian\inst{1}
 \and A.G.~Akhperjanian \inst{2}
 \and K.-M.~Aye \inst{3}
 \and A.R.~Bazer-Bachi \inst{4}
 \and M.~Beilicke \inst{5}
 \and W.~Benbow \inst{1}
 \and D.~Berge \inst{1}
 \and P.~Berghaus \inst{6}
 \and K.~Bernl\"ohr \inst{1,7}
 \and O.~Bolz \inst{1}
% ...
}

\offprints{J.A. Hinton, \email{Jim.Hinton@mpi-hd.mpg.de}}

\institute{
Max-Planck-Institut f\"ur Kernphysik, P.O. Box 103980, D 69029
Heidelberg, Germany
\and
 Yerevan Physics Institute, 2 Alikhanian Brothers St., 375036 Yerevan,
Armenia
\and
University of Durham, Department of Physics, South Road, Durham DH1 3LE,
U.K.
\and
Centre d'Etude Spatiale des Rayonnements, CNRS/UPS, 9 av. du Colonel Roche, BP
4346, F-31029 Toulouse Cedex 4, France
\and
Universit\"at Hamburg, Inst. f\"ur Experimentalphysik, Luruper Chaussee
149, D 22761 Hamburg, Germany
\and
Physique Corpusculaire et Cosmologie, IN2P3/CNRS, Coll{\`e}ge de France, 11 Place
Marcelin Berthelot, F-75231 Paris Cedex 05, France
\and
Institut f\"ur Physik, Humboldt-Universit\"at zu Berlin, Newtonstr. 15,
D 12489 Berlin, Germany
% ...
}

\date{Received ? / Accepted ?}

\abstract{

We report the detection of a point-like source of very high energy
(VHE) $\gamma$-rays coincident within $1'$ of Sgr A$^*$, obtained with
the H.E.S.S. array of Cherenkov telescopes. The $\gamma$-rays exhibit
a power-law energy spectrum with a spectral index of $-2.2 \pm 0.09
\pm 0.15$ and a flux above the 165~GeV threshold of $(1.82 \pm 0.22)
\cdot 10^{-7}$m$^{-2}$s$^{-1}$. The measured flux and spectrum differ
substantially from recent results reported 
in particular by the CANGAROO collaboration.
 
\keywords{gamma-rays: observations -- Galaxy: centre};
}

\maketitle

\section{Introduction}

The Galactic Centre (GC) region (Melia \& Falcke~\cite{Melia})
harbours a variety of potential sources of high-energy radiation
including the supermassive black hole Sgr A$^*$ of $ 2.6 \times 10^6
$~M$_\odot$ (see e.g. Sch\"odel et~al.~\cite{Schoedel}), which has been
identified as a faint source of X-rays (Baganoff
et~al.~\cite{Baganoff}) and infrared radiation (Genzel
et~al.~\cite{Genzel_flare}). Emission from Sgr A$^*$ is presumably
powered by the energy released in the accretion of stellar winds onto
the black hole (Melia~\cite{Melia92}; Yusef-Zadeh et~al.~\cite{Yusef}; Yuan
et~al.~\cite{yuan03}).  

High~(Mayer-Hasselwander et al.~\cite{egret}) and 
very high (Tsuchiya et al.~\cite{CANGAROO}; Kosack et al.~\cite{VERITAS})
energy $\gamma$-ray emission have also been detected from 
the GC region. The $\gamma$-radiation could result from
acceleration of electrons or protons in shocks in these winds, in the
accretion flow or in nearby supernova remnants, followed by
interactions of accelerated particles with ambient matter or
radiation.  Alternative mechanisms include the annihilation of dark
matter particles accumulating at the GC (Bergstr\"om et~al.~\cite{DM1};
Ellis et~al.~\cite{Ellis}; Gnedin \& Primack~\cite{Primack}) or
curvature radiation of protons near the black hole (Levinson~\cite{Levinson}).

\section{Observations and Results}

The observations presented here were obtained in Summer 2003 with the
High Energy Stereoscopic System (H.E.S.S.), consisting of four imaging
atmospheric Cherenkov telescopes (Hofmann~\cite{HESS}; Bernl\"ohr
et~al.~\cite{optics}; Vincent et~al.~\cite{camera}) in Namibia, at
$23^\circ 16'$~S $16^\circ 30'$~E.  At this time, two of the four
telescopes were operational, the other two being under
construction. During the first phase of the measurements (June 6 to
July 7, 2003), the telescopes were operated independently and images
were combined offline using GPS time stamps (4.7~h on source,
`June/July' data set).  In the second phase (July 22 to August 29,
2003), a hardware coincidence
required shower images simultaneously in both telescopes (11.8~h on
source, `July/August' data set).  The resulting background suppression
allowed us to lower the telescope trigger thresholds, yielding a
post-cuts energy threshold of 165~GeV (for typical Sgr A$^*$ zenith
angles of $20^\circ$) as compared to 255~GeV for the `June/July' data
set.

Shower images are parametrised by their centres of gravity and second
moments, followed by the stereoscopic reconstruction of shower
geometry, providing an angular resolution of
$\approx 0.1^\circ$
for individual $\gamma$-rays. $\gamma$-ray candidates are selected based on
the shape of shower images, allowing effective suppression of
cosmic-ray 
showers.  The $\gamma$-ray energy is estimated from
the image intensity and the reconstructed shower geometry, with a
typical resolution of 15-20\%.

The GC region is characterised by high night-sky brightness (NSB),
varying across the field of view and potentially interfering with
image reconstruction.  Simulations of a range of NSB levels show,
however, that the stereoscopic reconstruction is insensitive to this
feature, resulting in variations of the measured flux and spectrum
that are well within the systematic errors quoted here.

The performance and stability of H.E.S.S. have been confirmed by 
observations of the Crab Nebula (a standard candle in $\gamma$-ray
astronomy). The absolute calibration of the instrument has been verified using muon
images (Leroy et al.~\cite{Muon}) which provide a measurement of the absolute 
photon detection efficiency, and by the measured cosmic ray detection 
rates (Funk et al.~\cite{Trigger}), which are in excellent agreement 
with simulations. 

\begin{figure}
\centering
\includegraphics[height=7.5cm]{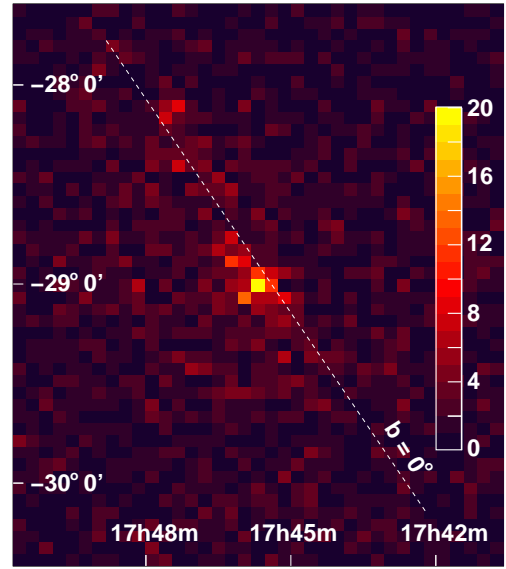}
\caption{Angular distribution of $\gamma$-ray candidates 
for a 3$^\circ$ field of view centred on Sgr~A$^*$. 
Both data sets ('June/July' and 'July/August') are combined,
employing tight cuts 
to reduce the level of background. 
The significance of the feature extending along the Galactic Plane is 
under investigation.}
\label{fig_locationA}
\end{figure}

\begin{figure}
\centering
\includegraphics[height=7.5cm]{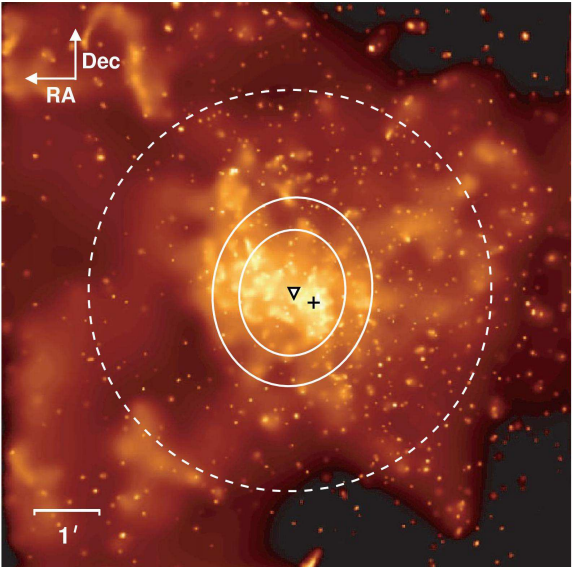}
\caption{Centre of gravity of the VHE signal (triangle), 
superimposed on a 8.5$'$ by 8.5$'$ Chandra X-ray map (Muno et al.~\cite{chandra_map})
of the GC. The location of Sgr~A$^*$
is indicated by a cross. The contour lines indicate the 68\%
and 95\% confidence regions for the source position, taking into 
account systematic pointing errors of 20$''$.
The white dashed line gives the 95\% confidence level upper limit
on the rms source size. The resolution for individual VHE photons 
- as opposed to the precision for the centre of the VHE signal - is
5.8$'$ (50\% containment radius).}
\label{fig_locationB}
\end{figure}

Fig.~\ref{fig_locationA} shows the distribution of $\gamma$-ray
candidates for a $3^\circ$ window around Sgr~A$^*$.  A clear excess of
events in the Sgr~A$^*$ region is observed.  Here, tight $\gamma$-ray
selection cuts are applied to minimise background at the expense of
$\gamma$-ray efficiency. For the analysis of the flux and spectrum of
the central point source, looser cuts are used which reject 96\% of
the cosmic-ray background and retain 50\% of the $\gamma$-rays. Using
a ring around the assumed source location to estimate background, we
find -- with loose cuts -- a 6.1~$\sigma$ excess in the `June/July'
data set and a 9.2~$\sigma$ excess in the `July/August' data set, both
centred on Sgr~A$^*$.  The $\gamma$-ray excess is located at RA
$17^h45^m41.3^s \pm 2.0^s$ , Dec $-29^\circ 0'22'' \pm 32''$, or
$l=359^\circ 56'53''$ , $b=-0^\circ 2'57''$, within $14 \pm 30 ''$ in
$b$ and $12 \pm 30 ''$ in $l$ from Sgr~A$^*$
(Fig.~\ref{fig_locationB}). There is no evidence in our data for an
energy dependence of this position. 
A conservative pointing error of less
than $20''$ in RA and Dec has been estimated using stars (Gillessen et
al.~\cite{Gillessen}), and verified by reconstructing the location of
known VHE sources such as the Crab Nebula and the AGN PKS 2155-304.

\begin{figure}
\centering
\includegraphics[width=8.5cm]{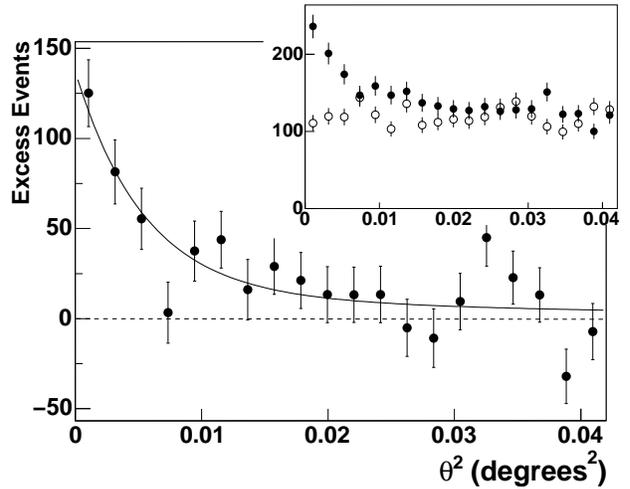}
\caption{Angular distribution of VHE $\gamma$-rays relative to the location
of Sgr~A$^*$. Inset: distributions in $\theta^2$ where $\theta$ is the
angle between the $\gamma$-ray direction and Sgr~A$^*$; a uniform background
results in a flat distribution in $\theta^2$. Full points: signal region;
open points: background region. The main figure shows 
background-subtracted excess counts. The solid line indicates
 the distribution expected for a point source of $\gamma$-rays at the 
position of Sgr~A$^*$.}
\label{fig_angular}
\end{figure}

Given the high density of potential sources over the central square
degree of the Galaxy, an important question is whether the VHE
$\gamma$-ray signal 
shows signs of source extension. Fig.~\ref{fig_angular} shows the angular
distribution of detected $\gamma$-rays relative to Sgr A$^*$.  The width
and shape of this distribution are consistent with point source 
simulations. These simulations have been verified using the strong
signals from the Crab Nebula and PKS 2155-304.  Assuming a Gaussian
distribution of source brightness, $\rho \propto \exp(-\theta^{2}/2
\sigma_{source}^2)$, we find an upper limit $\sigma_{source} < 3'$ for
the source size (95\% CL), corresponding to $< 7$~pc at the distance
of the GC (dashed white line in Fig.~\ref{fig_locationB}).  The
apparent point-like nature of the central source does not exclude the
possibility of non-azimuthally symmetric tails in the emission.

The measured energy spectrum is shown in Fig.~\ref{fig_spectrum}.
Data are fit by a power-law, 
$F(E) = F_0 E_{\mathrm{TeV}}^{-\alpha}$,
with a spectral index $\alpha = 2.21 \pm 0.09$ and $F_0 = (2.50 \pm
0.21) \times 10^{-8}$m$^{-2}$s$^{-1}$TeV$^{-1}$ for the `July/August'
data set (full circles), with a $\chi^2$/d.o.f. of 0.6.  The flux
above the 165~GeV threshold is $(1.82 \pm 0.22) \times
10^{-7}$m$^{-2}$s$^{-1}$, equivalent to 5\% of the Crab Nebula flux at
this threshold.  The smaller `June/July' data set gives
consistent results ($\alpha = 2.11 \pm 0.19$ and $F_0 = (2.76 \pm 0.33)
\times 10^{-8}$m$^{-2}$s$^{-1}$TeV$^{-1}$).  We estimate systematic
errors of $\Delta \alpha \approx 0.15$ and $\Delta F/F \approx 25\% $,
with the latter mainly governed by the precision of the energy
calibration of the instrument.  The energy reconstruction and flux
determination have been tested with the Crab Nebula; we reconstruct a
power law with index $\alpha=2.63\pm0.04$ and a flux above 1 TeV of
$(1.98\pm0.08)\times 10^{-7}$m$^{-2}$s$^{-1}$ for the Crab, 
in very good agreement with previous measurements (see Aharonian et~al.~\cite{Aharonian} 
and references therein). Fitting the GC $\gamma$-ray spectrum as a 
power law with an exponential cutoff, we find a lower limit for the 
cutoff energy of 4~TeV.  Within statistics, there are no indications 
for time variability of the GC signal.

\section{Discussion and Conclusions}

\begin{figure}
\centering
\includegraphics[width=8.5cm]{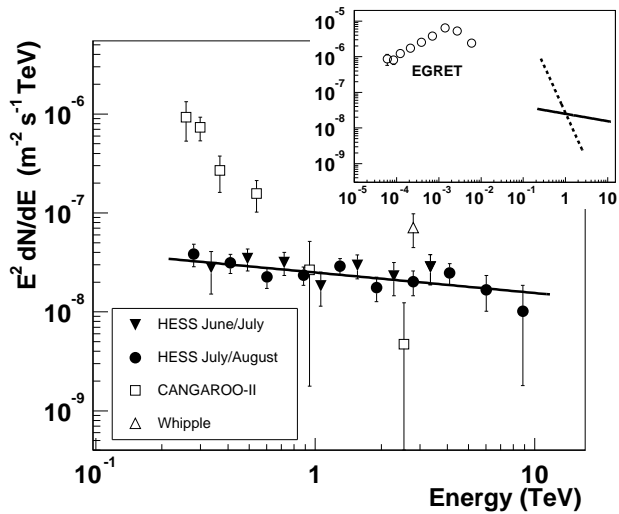}
\caption{Energy spectrum $E^2 dN/dE$ of $\gamma$-rays from the Galactic Centre.
Full circles: H.E.S.S. `July/August 2003' data set. Full triangles: H.E.S.S. 
`June/July 2003' data set.
The line indicates a power-law fit to the `July/August'
spectrum. Open squares: CANGAROO-II spectrum from Summer 2001 and 2002 
(Tsuchiya et al.~\cite{CANGAROO}). 
Open triangle: Whipple flux from 1995 through 2003
(Kosack et al.~\cite{VERITAS}), converted to a differential flux at the peak detection
energy assuming a Crab-like spectrum. 
The inset shows the EGRET 
flux from 1991 to 1996 (Mayer-Hasselwander et al.~\cite{egret}) (circles) compared to fits
to the CANGAROO-II (dashed line) and H.E.S.S. (solid line) spectra. 
Due to the poor angular resolution of EGRET ($1^\circ$) the flux shown may 
include other sources.
}
\label{fig_spectrum}
\end{figure}

The CANGAROO collaboration recently reported the detection of sub-TeV
gamma rays from within $0.1^\circ$ of the GC based on 67~h of
(on-source) data taken in July 2001 and July/August 2002
(Tsuchiya et al.~\cite{CANGAROO}). The reported spectrum is very steep, $F(E) \propto
E^{-4.6 \pm 0.5}$.  The rather hard H.E.S.S. spectra are obviously not
consistent with the steep spectrum obtained with CANGAROO-II
(Fig.~\ref{fig_spectrum}); the large flux at low energies implied by
the CANGAROO result would have been detected with H.E.S.S. in a matter
of minutes.  At higher energies, above 2.8~TeV, a marginal detection
with a significance of 3.7~$\sigma$ resulting from 26~h of
large-zenith-angle observations in the years 1995 through 2003 was
reported by the Whipple collaboration (Kosack et al.~\cite{VERITAS}),
consistent with
Sgr~A$^*$ within the $15'$ 95\% C.L. error circle.  The Whipple
flux is a factor 3 above that implied by our spectra.  Taking all data
at face value, one would conclude that the source underwent
significant changes over the timescale of one year (2002 to 2003).
However, this seems unlikely since none of the individual experiments detects
significant variability.  Implications of the CANGAROO and Whipple
data are discussed in Hooper et al.~(\cite{silk}).

At lower energies around 100~MeV, the EGRET instrument detected a
strong excess from the central part of the Galaxy (Mayer-Hasselwander et al.~\cite{egret}), 
consistent within its error circle with Sgr~A$^*$, but with angular
resolution of $1^\circ$ also covering other potential sources
(and a solid angle $\sim$100 times larger than the emission 
region seen by H.E.S.S.). In the
analysis of a high-energy sub-sample of EGRET data (Hooper \& Dingus~\cite{egret_reana}),
the source was found to be offset from the GC by $0.21^\circ$, 
excluding Sgr~A$^*$ at 99.9\% C.L.

Models for the wide-band spectra of Sgr~A$^*$ include Advection
Dominated Accretion Flow (ADAF) models (Narayan et al.~\cite{Narayan98};
Yuan et al.~\cite{yuan03}),
possibly combined with a jet extracting energy from the accretion disk
(Yuan et al.~\cite{Falcke}).  Shocks in the accretion flow (Markoff
et al.~\cite{Markoff}) or in the jet could accelerate particles. 
$\gamma$-rays are generated in proton
interactions, but predicted spectra tend to fall off rapidly in the
TeV region (Markoff et al.~\cite{Markoff}).

Another source of VHE $\gamma$-rays should be diffuse emission from the
entire central region, in which case year time-scale variability should
not occur. $\gamma$-rays may result from interactions of  
accelerated protons and nuclei (Fatuzzo \& Melia~\cite{fatuzzo03})
with the ambient matter with a density
as large as $n=10^3\,\rm cm^{-3}$ (Maeda et al.~\cite{Maeda}).  
Only modest overall energy, 
$W_{\rm p} \simeq 5\,\times\,10^{47} (10^{3}\,\rm cm^{-3} / n) \ \rm erg$,
in TeV protons is needed to explain the observed $\gamma$-ray flux from
this region (the 1-10~TeV luminosity is $\sim \, 10^{35}$ ergs/s).  
An obvious candidate for the proton accelerator could be
the young ($10^4$ yr) and unusually powerful (total explosion energy
$\simeq 4 \times 10^{52} \ \rm erg$) supernova remnant Sgr A East
(Maeda et al.~\cite{Maeda}).  The measured spectral index of TeV emission, 
$\alpha \approx 2.2$, is close to the spectrum of shock-accelerated particles.
For a $10^4$~yr source age, the modest source extension and the hard
spectrum imply that particle diffusion in the central region proceeds
much slower compared to the diffusion in the Galactic Disk.
While detailed modelling remains to be done, estimates show that for
magnetic fields up to $\sim$1~mG the X-ray and radio emission resulting from
secondary electrons generated in interactions of such accelerated
protons are below the measured diffuse luminosities integrated over
Sgr~A East. This consistency criterion is, however, much more
challenging if one considers Sgr~A* as the source, with its
significantly larger magnetic fields and lower (quiescent-state) X-ray
flux. In either case, the explanation of the EGRET flux (Fig.~\ref{fig_spectrum})
requires a second source component with a cutoff close to the highest EGRET
energies, located well within the H.E.S.S. field of view (and therefore
excluded as a strong TeV source), but not necessarily coincident with
Sgr A East or Sgr A*. We note that Tsuchiya et al.~(\cite{CANGAROO}) 
describe the EGRET and CANGAROO fluxes jointly in a model of a diffuse 
proton flux with a spectral cutoff of a few TeV, interacting with ambient gas. 
In such a model, one would, however, not expect fast variability.

Alternative mechanisms invoke the hypothetical annihilation 
of super-symmetric dark matter particles 
(Bergstr\"om et~al.~\cite{DM1};
Ellis et~al~\cite{Ellis}; Gnedin \& Primack~\cite{Primack})
or curvature radiation of protons in the vicinity of the central  
super-massive black hole (Levinson~\cite{Levinson}).

The spectrum of $\gamma$-rays from hypothetical annihilation of
neutralinos of mass $M_\chi$ consists of a $\gamma$-ray continuum and two
lines at $E = M_\chi$ and $E = M_\chi (1-m_Z^2/4 M_\chi^2)$. The
continuum spectra generated by the DarkSusy program (Gondolo et al.~\cite{DarkSusy})
are well approximated by $F \sim E^{-\alpha} e^{-(E/M_{\mathrm{cut}})}$ with
$\alpha = 2.2...2.4$ and $M_{\mathrm{cut}} = 0.15...0.3 M_\chi$ depending on
the annihilation channel
\footnote{Bergstr\"om et~al.~(\cite{DM1}) use $\alpha = 1.5$ and $M_{\mathrm{cut}} = 0.13
M_\chi$ to describe spectra for $E < 0.5 M_\chi$, but with larger
$\alpha$ values a better fit up to $E = 0.9 M_\chi$ is
obtained.}. Assuming that the observed $\gamma$-rays represent a
continuum annihilation spectrum, the lower limit of 4~TeV on the
cutoff implies $M_\chi > 12$~TeV, a range which is presently 
disfavoured due to
particle physics and cosmology arguments (Ellis et al.~\cite{EllisWMAP}).
Supersymmetric dark matter annihilation as the main source of the
observed $\gamma$-rays is therefore unlikely, but not excluded.

The spectrum of the proton curvature radiation (Levinson~\cite{Levinson})
depends, to a large extent, on the configuration of magnetic fields
near the gravitational radius of the black hole, and detailed
predictions are lacking; as a characteristic feature, one would expect
time variability.  Further observations of the GC region are a high
priority for H.E.S.S. in the near future.

\begin{acknowledgements}

The support of the Namibian authorities and of the University of Namibia
in facilitating the construction and operation of H.E.S.S. is gratefully
acknowledged, as is the support by the German Ministry for Education and
Research (BMBF), the Max Planck Society, the French Ministry for Research,
the CNRS-IN2P3 and the Astroparticle Interdisciplinary Programme of the
CNRS, the U.K. Particle Physics and Astronomy Research Council (PPARC),
the IPNP of the Charles University, the South African Department of
Science and Technology and National Research Foundation, and by the
University of Namibia. We appreciate the excellent work of the technical
support staff in Berlin, Durham, Hamburg, Heidelberg, Palaiseau, Paris,
Saclay, and in Namibia in the construction and operation of the
equipment.

\end{acknowledgements}

\end{document}